\documentclass[useAMS]{mn2e} 
\usepackage{times}

\input{epsf}

\title[X-ray/$\gamma$-ray spectrum of XTE J1550--564] 
{The X-ray/$\gamma$-ray spectrum of XTE J1550--564 in the very high 
state}

\author[M. Gierli\'nski, C. Done]
{Marek~Gierli\'nski$^{1,2}$ and Chris Done$^1$\\
$^1$Department of Physics, University of Durham, South Road, Durham DH1 3LE, 
UK\\ 
$^2$Obserwatorium Astronomiczne Uniwersytetu Jagiello{\'n}skiego, 30-244 
Krak{\'o}w, Orla 171, Poland}

\date{Submitted to MNRAS}
\pagerange{\pageref{firstpage}--\pageref{lastpage}}
\pubyear{2002}

\begin{document}

\def\lh{$\ell_{\rm h}$}
\def\lth{$\ell_{\rm th}$}
\def\lnth{$\ell_{\rm nth}$}
\def\ls{$\ell_{\rm s}$}
\def\lhs{$\ell_{\rm h}/\ell_{\rm s}$}
\def\lnh{$\ell_{\rm nth}/\ell_{\rm h}$}

\def\aap{A\&A}
\def\apj{ApJ}
\def\apjl{ApJ}
\def\mnras{MNRAS}

\topmargin = -0.5cm

\maketitle

\label{firstpage}

\begin{abstract}

We fit the broad-band X/$\gamma$-ray spectrum (0.8--1000~keV) of 
the accreting black hole XTE J1550--564 in the very high state. 
The quasi-simultaneous data from {\it ASCA}, {\it RXTE\/} and 
OSSE show that the disc is strongly Comptonized, with a high 
energy tail extending out to several hundreds keV. However, 
inverse Compton scattering by a purely thermal or purely 
power-law electron distribution cannot explain the observed 
spectrum. Instead the data require a hybrid distribution, with 
both thermal and non-thermal electrons scattering the disc 
photons. This is very similar to the electron distribution 
inferred for other high and very high state black hole binaries, 
showing that it is a generic feature of high mass accretion rate 
black holes.

\end{abstract}



\section{Introduction}
\label{sec:introduction}

Galactic black hole candidates show variety of spectra at X-ray 
and $\gamma$-ray energies. These can be roughly described in 
terms of a soft, quasi-thermal component from the accretion disc 
together with a power-law-like tail extending to much higher 
energies. Five `canonical' spectral states have been identified 
historically, based on the luminosity and spectral shape (see 
e.g.\ Esin, McClintock \& Narayan 1997 and references therein). 
The {\em very high state\/} is generally seen at the highest 
luminosities, where the soft thermal component (with temperature 
of 0.5--1 keV) and power-law tail (photon spectral index 
$\Gamma$ = 2--3) have comparable power. In the {\em high 
state\/} the hard tail is much weaker or sometimes even not 
present. The {\em intermediate state\/} is rather like the very 
high state, but at lower disc temperature and luminosity 
(Belloni et al.~1996; M{\'e}ndez \& van der Klis 1997), while 
the {\em low state\/} is somewhat different, being characterised 
by a hard ($\Gamma < 2$) power-law-like spectrum with a very 
weak, low temperature disc component (Ba{\l}uci{\'n}ska-Church 
et al.~1995; di Salvo et al.~2001). At very low luminosities, 
transient sources go into the {\em quiescent state}, which is a 
very low luminosity version of the {\em low state\/} (Kong et 
al.~2000; Sutaria et al.~2002). The high and very high states 
are often described as soft states, while the low state is 
commonly referred to as a hard state. Persistent sources like 
Cyg X-1 and LMC X-1 spend most of their time in one state, 
however transients like Nova Muscae can progress through a 
sequence of all five states during an outburst (Ebisawa et 
al.~1994). See Done \& Gierli{\'n}ski (2003) for an overview of 
these spectral states.

The power-law-like tails extending to X/$\gamma$-ray energies 
are commonly explained by Compton up-scattering of the soft seed 
photons from the accretion disc by a population of high-energy 
electrons. In the low/hard state the observed hard spectra roll 
over at 100--200 keV, indicating that the electrons are mostly 
thermal (with typical temperatures of $\sim$ 100 keV) and have 
optical depth of order unity (e.g.\ Gierli{\'n}ski et al.~1997; 
Zdziarski et al.~1998; Frontera et al.~2001). By contrast, in 
the soft states the steeper tail extends unbroken to MeV 
energies (Grove et al.~1998), indicating that the electron 
energy distribution is predominantly non-thermal. More detailed 
spectral studies of the very high- and high-state spectra show 
that the tail is best described by Comptonization from a complex 
electron distribution, where there is both a low temperature 
($\sim 10$ keV) thermal electron distribution ({\.Z}ycki, Done 
\& Smith 2001; Kubota, Makishima \& Ebisawa 2001) together with 
non-thermal power-law electrons (Gierli{\'n}ski et al.~1999; 
Zdziarski et al.~2002). A two-component electron distribution 
could arise fairly naturally from non-thermal electron 
acceleration regions powered by magnetic reconnections above a 
disc. Low-energy electrons cool preferentially by Coulomb 
collisions leading to a thermal distribution while the 
high-energy electrons cool by Compton scattering, preserving a 
non-thermal distribution (Coppi 1999). Alternatively, the 
thermal and non-thermal electrons could be spatially distinct, 
e.g.\ magnetic reconnection above the disc can produce a 
non-thermal electron distribution, while overheating of the 
inner disc produces the thermal Comptonization (Kubota et 
al.~2001).

The soft and hard states are clearly distinct in their spectral 
properties (e.g. Zdziarski et al.~2002). Transition between the 
states is compatible with being driven by a changing accretion 
geometry, where the low state has an accretion disc only at 
large radii, with the inner disc being replaced by a hot, 
optically thin, geometrically thick flow (e.g.\ Poutanen, Krolik 
\& Ryde 1997). With increasing accretion rate the disc moves 
inwards increasing its temperature and the number of seed 
photons cooling the inner flow so the Comptonized spectrum 
becomes softer. Eventually, when the inner flow becomes 
optically thick it collapses, the inner disc replaces the hot 
flow and the source reaches one of the soft states (see e.g.\ 
Esin et al.~1997).

This review of spectral properties draws heavily from the very 
few broad bandpass X/$\gamma$-ray observations of black holes. 
To constrain both the disc and hard component requires 
simultaneous data over a broad energy range, ideally $\sim$ 
0.1--1000 keV. While both the Phoswich Detector System on board 
{\it BeppoSAX} and High Energy X-ray Timing Experiment (HEXTE) 
on board {\it Rossi X-ray Timing Explorer\/} ({\it RXTE}) can 
give data out to $\sim$ 100--200 keV, the Oriented Scintillation 
Spectrometer Experiment (OSSE) on the {\it Compton Gamma Ray 
Observatory\/} ({\it CGRO\/}) is unique in being able to extend 
the bandpass to higher energies. This is especially important in 
soft states, as there the hard component is known to extend to 
the highest energies. However, there are only very few 
broad-band observations of black hole candidates in these 
spectral states: GRS~1915+105 (Zdziarski et al.~2001; Ueda et 
al.~2002), GRO~J1655--40 (Tomsick et al.~1999), Cyg X-1 
(Gierli{\'n}ski et al.~1999; McConnell et al.~2002). However, 
both GRO~J1655--40 and GRS~1915+105 are thought to be extreme 
Kerr black holes due to their high accretion disc temperatures 
(e.g.\ Zhang, Cui \& Chen 1997), so there are {\it no\/} very 
high state broad-band spectra of `normal' black holes.

In this paper we remedy the situation by analysing the 
broad-band simultaneous {\it RXTE}/OSSE data of the X-ray black 
hole transient XTE J1550--564, and derive constraints on the 
electron distribution and accretion flow geometry. We show that 
the hard tail is remarkably similar to that seen in the extreme 
Kerr, microquasar black holes, so it is most probably a generic 
feature of high mass accretion rate flow onto a black hole 
rather than being associated with spin or radio jets. 

\section{The source}
\label{sec:source}

The X-ray nova XTE J1550--564 was discovered during its 
spectacular outburst in 1998 (Smith et al.~1998). Two weeks into 
the outburst it reached a peak flux of 6.8 crab ($1.6\times 
10^{-7}$ erg s$^{-1}$ cm$^{-2}$) in 2--10 keV energy range, 
making it the brightest X-ray transient observed so far by {\it 
RXTE}.  The X-ray (2--10 keV) lightcurve from All-Sky Monitor 
(ASM), is more complex than a fast rise, exponential decay, 
having a bright flare followed by two plateaus separated by a 
weak minimum. Such lightcurves are not uncommon (see e.g.\ Chen, 
Shrader \& Livio 1997), and are probably connected to enhanced 
mass transfer from the irradiated companion star (Esin, Lasota 
\& Hynes 2000). The complex X-ray spectral evolution during the 
outburst is shown in Fig.~\ref{fig:lightcurves} by comparing the 
2--10 keV lightcurve from ASM to the harder X-ray (20--100 keV) 
lightcurve from {\it CGRO\/} Burst And Transient Source 
Experiment (BATSE). More details of this spectral evolution are 
given in Sobczak et al.~(2000), Homan et al.~(2001), Wilson \& 
Done (2001) and Kubota et al.~(2003).

Observations of the optical companion (Orosz et al.~2002) led to 
an estimate of the compact object mass, 8.4--10.8 M$_\odot$, 
confirming its black hole nature. The inclination angle of the 
binary is quite high, $67^\circ$--$77.4^\circ$ (both estimates 
are given with $1\sigma$ errors, the $3\sigma$ ranges are 
6.8--15.6 M$_\odot$ and $55^\circ$--$79^\circ$, respectively). 
The distance is less well constrained by their data on the 
optical companion, with $D$ = 3.0--7.6 kpc (best fit value of 
5.3~kpc). Throughout this paper we assume $M$ = 10M$_\odot$, $i 
= 70^\circ$ and $D$ = 5.3 kpc.

\section{The data}
\label{sec:data}

\begin{table*}

\caption{Log of observations.}

\label{tab:obslog}
\begin{tabular}{lccccc}
\hline
Instrument & Detectors & Observation ID & Time (MJD) & Live time (s) & Count rate (cnt s$^{-1}$)\\
\hline
{\it ASCA\/} GIS & 2 & 15606010 & 51079.964--51080.500 & 25156 & 93.30$\pm$0.06\\
{\it RXTE\/} PCA & 01, layer 1 & 30191-01-12-00 ... -23-00 & 51082.003--51091.790 & 45664 & 5891$\pm$6\\
{\it RXTE\/} HEXTE & 0 & 30191-01-12-00 ... -23-00 & 51082.003--51091.790 & 14442 & 287.8$\pm$0.2\\
{\it CGRO\/} OSSE & 1234 & 729.5 & 51081.640--51092.561 & 496313 & 12.2$\pm$0.2\\
\hline
\end{tabular}
\end{table*}

\begin{figure}
\begin{center} 
\leavevmode 
\epsfxsize=8cm \epsfbox{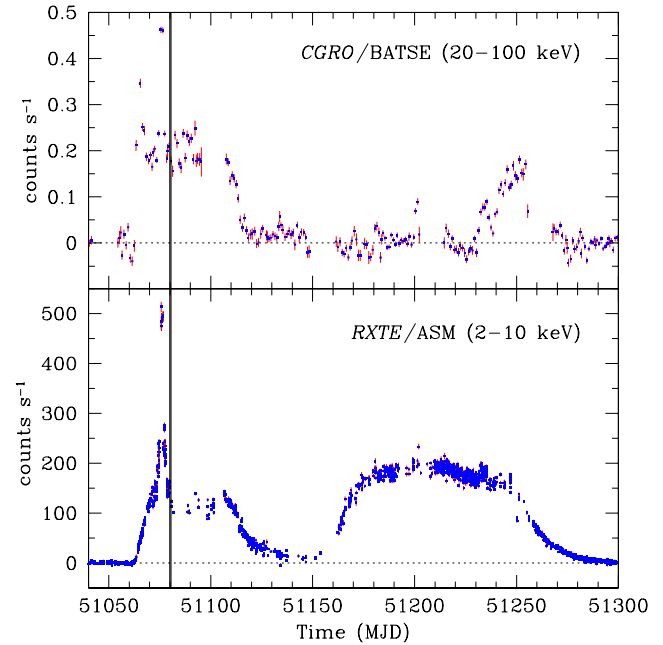} 
\end{center} 

\caption{BATSE and ASM lightcurves of 1998 outburst of XTE 
J1550--564. Vertical line marks the OSSE viewing period 
729.5, analysed in this paper together with simultaneous {\it RXTE\/} 
observations. The {\it ASCA\/} observations was made one day before 
the OSSE viewing period 729.5 had started.}

\label{fig:lightcurves} 
\end{figure}

In this paper we analyse the contemporaneous X/$\gamma$-ray 
spectra of XTE J1550--564 from {\it ASCA}, {\it RXTE\/} and OSSE 
in a period from 1998 September 23 to 1998 October 6 (MJD 
51080--51092) i.e. in the first part of the outburst, in the 
plateau just after the strong flare. There are {\it RXTE\/} 
observations over this entire period, while the {\it ASCA\/} 
data are limited to a snapshot at the start (September 23/24) 
which does not quite overlap with the {\it OSSE\/} data (between 
September 25 and October 6). The times of our data are shown on 
the soft and hard X-ray lightcurves of the outburst in 
Fig.~\ref{fig:lightcurves} while a detailed observation log is 
presented in Table \ref{tab:obslog}.
 
We used the standard product {\it ASCA\/} Gas Imaging 
Spectrometer (GIS) spectrum from detector 2, with response 
matrix version 4.0, corrected for dead time (Makishima et 
al.~1996). Since the source is very bright we fit the GIS2 
spectrum alone to avoid possible cross-calibration uncertainties 
between GIS2 and GIS3. There is a simultaneous {\it RXTE\/} 
pointing (observation number 30191-01-10-00), but we treat the 
{\it ASCA\/} data separately to avoid the significant 
cross-calibration issues between {\it RXTE\/} and {\it ASCA\/} 
(e.g. Done, Madejski, {\.Z}ycki 2000).


The {\it OSSE\/} data for viewing period 729.5 extends between 
September 25 and October 6 (Fig.~\ref{fig:lightcurves}). We use 
a high-level product spectrum extracted for detectors 1, 2, 3 
and 4, for the whole viewing period. Due to rather large OSSE 
field of view ($11^\circ \times 3.8^\circ$; Johnson et al.~1993) 
there is always a danger of a background source contaminating 
the spectrum. Therefore, we compare the OSSE spectrum with the 
simultaneous HEXTE spectrum. Since HEXTE has much smaller field 
of view ($\sim 1^\circ$; Rothschild et al.~1998), contamination 
by a background source is much less probable. The excellent 
agreement between the HEXTE and OSSE spectra (see Section 5.2) 
make contamination by a background source rather unlikely.


\begin{figure}
\begin{center} 
\leavevmode 
\epsfxsize=7cm \epsfbox{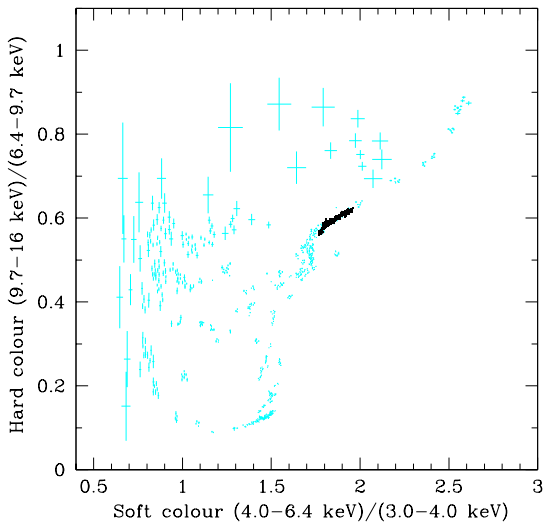} 
\end{center} 

\caption{{\it RXTE}/PCA colour-colour diagram of the 1998 outburst of 
XTE J1550--564. The colours were corrected for the PCA gain changes 
(by the same method as in Gierli{\'n}ski \& Done 2002) and rebinned 
for clarity. Data points in black denote the {\it RXTE\/} 
observations simultaneous with OSSE viewing period 729.5.}

\label{fig:colcol} 
\end{figure}


{\it RXTE\/} observed the source 15 times during OSSE viewing 
period 729.5, with observation archival numbers (obsids) between 
30191-01-12-00 and 30191-01-23-00. To assess the source X-ray 
variability over this 10 day period we create a PCA 
colour-colour diagram, using a method described by 
Gierli{\'n}ski \& Done (2002). The diagram, shown in 
Fig.~\ref{fig:colcol}, contains all the PCA data collected 
during the 1998/1999 outburst, and observations simultaneous 
with OSSE are marked in black. There is a shift in colours 
during this period, though it is small compared to the spectral 
variation seen during the whole outburst. This particular part 
of the colour-colour diagram corresponds to the very high 
spectral state (Done \& Gierli{\'n}ski 2003).

We extract the PCA and HEXTE spectra using {\sc ftools} 5.2 for 
each of the aforementioned obsids, and also create a single 
spectrum for each instrument averaged over the whole OSSE 
viewing period. For the PCA, we extract the spectra from 
detectors 0 and 1, top layer only (see Wilson \& Done 2001), 
correcting for the photon pile-up (Tomsick et al.~1999). For 
HEXTE we extract spectra from cluster 0 only.


\section{Spectral models}
\label{sec:models}

As outlined in the introduction, the X-ray spectrum is probably 
formed by Compton scattering of the accretion disc seed photons. 
While an exponentially cutoff power law is often used as an 
approximation to a thermally Comptonized spectrum, this becomes very 
inaccurate if the observed bandpass approaches the typical seed 
photon or electron energies (e.g.\ Done, {\.Z}ycki \& Smith 2002; 
Zdziarski et al.~1998). Since we have an extremely broad bandpass we 
use instead the best available Comptonization models, which properly 
calculate the shape of the high-energy rollover for thermal 
electrons, and include the low-energy cutoff close to the seed photon 
energies. There are currently two of these Comptonization codes 
available for spectral fitting, {\sc compps} (Poutanen \& Svensson 
1996) and {\sc eqpair} (Coppi 1999). Both of these can calculate the 
spectra from thermal as well as non-thermal Comptonization.

The {\sc compps} code computes Comptonization in various 
specific geometries (we assume spherical) from a given electron 
distribution of total optical depth $\tau$. This is split 
between a thermal distribution (Maxwellian, with temperature 
$T_e$) which is truncated at some Lorentz factor $\gamma_{\rm 
min}$, above which it has a non-thermal [power law, $N(\gamma) 
\propto \gamma^{-p}$] distribution up to Lorentz factor 
$\gamma_{\rm max}$.

The {\sc eqpair} code is rather different in that it computes the 
electron distribution self-consistently, assuming that the electrons 
are injected with a power-law distribution of index $\Gamma_{\rm 
inj}$.  The rate of Compton cooling is larger at higher energies, so 
the equilibrium distribution is softer than the injected one with 
index $p = \Gamma_{\rm inj} + 1$ (e.g.\ Blumenthal \& Gould 1970). 
The electrons can also thermalize through Coulomb collisions, 
producing a hybrid, thermal/non-thermal distribution. However, the 
code also allows some fraction of the total power to be used to heat 
the thermal distribution directly. The microscopic processes are 
calculated in the plasma cloud of assumed optical depth of ionization 
electrons, $\tau_{\rm ion}$. The total optical depth, $\tau = 
\tau_{\rm ion} + \tau_{\rm pair}$, is calculated self-consistently, 
including the additional optical depth in electron-positron pairs, 
$\tau_{\rm pair}$, produced by photon-photon collisions. The plasma 
properties depend on its compactness \begin{equation}\ell\equiv 
{\sigma_{\rm T} \over m_{\rm e} c^3} {{\cal L} \over {\cal 
R}},\end{equation} where ${\cal L}$ is the power supplied to the 
plasma, ${\cal R}$ is its characteristic size, $\sigma_{\rm T}$ is 
the Thompson cross section and $m_e$ is the electron mass. The power 
supplied to the electrons is parameterized as a hard compactness, 
\lh, while the power in soft seed photons is given as a soft 
compactness, \ls. The hard compactness is split into non-thermal 
electron acceleration and direct thermal heating, denoted \lnth\ and 
\lth, respectively, where \lh\/ = \lnth + \lth. The electron 
temperature, $T_e$, is calculated self-consistently from the balance 
between cooling and heating. We note that even when the whole power 
is supplied to non-thermal electrons (i.e.\ \lth\/ = 0), the Coulomb 
cooling can dominate at lower energies over Compton cooling, so the 
electrons will form a hybrid distribution. The model parameters are 
\ls, \lhs, \lnh, $\Gamma_{\rm inj}$ and $\tau_{\rm ion}$.

Both models include the reflected continuum computed from the 
incident continuum using the Green's functions of Magdziarz \& 
Zdziarski (1995). This is relativistically smeared (Fabian et 
al.~1989), but does not include the self-consistently produced 
iron K$\alpha$ line, so we include this separately using the 
{\sc diskline} model (Fabian et al.~1989), with relativistic 
smearing set equal to that of the reflected continuum. These are 
parameterized by the solid angle of the reflecting medium, 
$\Omega$, its ionization, $\xi$, and the inner disc radius, 
$R_{\rm ref}$. The rest-frame energy of the line is denoted as 
$E_{\rm line}$ and its equivalent width as $EW$.

For purely thermal Comptonization we also use a more approximate (but 
faster) {\sc thcomp} code (Zdziarski, Johnson \& Magdziarz 1996), 
which is based on Kompaneets (1956) equation. This assumes that the 
photons diffuse in both time and space (Sunyaev \& Titarchuk 1980), 
which becomes inaccurate for temperatures above $\sim$ 50 keV, as 
such photons lose a large fraction of their energy in Compton 
down-scattering. 

In all of these models we assume that the seed photons for 
Comptonization have a multi-colour disc blackbody (model {\sc 
diskbb}; Mitsuda et al.~1984) distribution, with temperature 
$T_{\rm seed} = T_{\rm diskbb}$. We allow for additional disc 
photons which do not pass through the Comptonizing region, e.g. 
corresponding to a fraction of the disc not covered by a patchy 
corona. For calculating inner disc radius we take into account 
all disc photons, both seed and non-seed. In figures we plot 
only the disc photons which reach the observer unscattered.


\section{Results}
\label{sec:results}

For spectral fitting we use the X-ray spectral fitting package 
{\sc xspec} version 11 (Arnaud 1996), with our own 
implementation of the Comptonization models described above. The 
error of each model parameter is given for a 90 per cent 
confidence interval ($\Delta\chi^2=2.7$). We use the 0.9--10 keV 
GIS data, 3--20 keV PCA data, 20--150 keV HEXTE data and 
50--1000 keV OSSE data. The relative normalizations of the PCA, 
HEXTE and OSSE are uncertain, so we allow them to be free 
parameters in simultaneous spectral fits. We base fluxes on the 
PCA normalization.

\subsection{\it ASCA}
\label{sec:asca}

The near-simultaneous {\it ASCA\/} observation gives us an 
excellent opportunity to study the soft X-ray part of the XTE 
J1550--564 spectrum, not covered by {\it RXTE\/} and OSSE. From 
the PCA data we know that the X-ray spectrum did not change much 
between {\it ASCA\/} and OSSE observations, so we can apply the 
seed photon temperature, $T_{\rm seed}$, and Galactic 
absorption, $N_H$, obtained from {\it ASCA\/} to the {\it 
RXTE}/OSSE spectral fits.

We use all three Comptonization models to check that the results 
do not depend on the details of the models used. The {\it 
ASCA\/} data are not sensitive to the details of the high energy 
spectrum, so we assume that the Comptonizing electrons are 
thermal with electron temperature fixed at $kT_e$ = 30 keV ({\sc 
compps} and {\sc thcomp}) or equivalently $\tau_{\rm ion} = 
2$ for {\sc eqpair}. 

Our model consists of this thermal Comptonization (together with 
its reflected continuum and associated iron line emission), and 
a multicolour disc blackbody.  All three models for the 
Comptonized continuum give almost identical values of $N_H$ and 
$T_{\rm seed}$ (Table \ref{tab:asca_fits}). We check that our 
choice of $T_e$ (or equivalently $\tau_{\rm ion}$) did not affect 
these values, and even replacing the disc spectrum by a 
single-temperature blackbody gives only a slightly lower column 
of about $6.0\times 10^{21}$ cm$^{-2}$. 

This result is not very different from previous findings. Jain 
et al.~(1999) found $N_H \approx 9\times10^{21}$ cm$^{-2}$ from 
optical reddening in the direction of XTE J1550--564. Kong et 
al.~(2002) fit the {\it Chandra\/} spectra of XTE J1550--564 in 
quiescence with various models and found $N_H$ between 3 
(blackbody) and $9\times10^{21}$ cm$^{-2}$ (power law). Thought 
it not clear what spectral shape should we expect from a black 
hole binary in a quiescence, Tomsick, Corbel \& Kaaret (2001) 
point out that the blackbody model leads to rather unphysical 
size of the source, $\sim$ 0.1 km. Raymond-Smith, bremsstrahlung 
and power law models with {\it Chandra\/} data all give $N_H$ 
consistent with our result. We note that a much larger column of 
$2\times10^{22}$ cm$^{-2}$ found by Sobczak et al.~(2000) from 
the PCA fits is due to lack of sensitivity of the PCA below 
$\sim$ 3 keV. And indeed, when we allow $N_H$ to be a free 
parameter in our {\it RXTE}/OSSE fit (see below, Section 
\ref{sec:rxte_osse}), it yields $N_H \approx 2.5\times10^{22}$ 
cm$^{-2}$.

Strongly ionized reflection is significantly detected in all 
these models (see also Wilson \& Done 2001), but rather 
surprisingly the amount of relativistic smearing is small. From 
all three models we find very high inner disc radius, typically 
$R_{\rm ref} > 200 R_g$ (where $R_g \equiv GM/c^2$ is the 
gravitational radius). However, closer investigation of the 
residuals shows that this is driven by narrow features in the 
spectrum around the iron line energy. A disc which extends much 
further into the gravitational potential is consistent with the 
data if there is a narrow absorption line at $\sim 7.2$ keV with 
40 eV equivalent width.  Such narrow features are seen from 
other X-ray binary systems viewed at high inclination, as GRO 
J1655--40 (Ueda et al.~1998) or GRS 1915+105 (Kotani et 
al.~2000), where it is interpreted as resonance absorption from 
iron, but the line energy we obtain in XTE J1550--564 is 
slightly too high. A P Cygni line profile, as seen in Cir X-1 
(Brandt \& Schulz 2000), gives an equally good fit to the XTE 
J1550--564 data, but with emission and absorption line energies 
($\sim 6.9$ and $7.0$ keV, respectively) which are physically 
consistent with the resonant transition in Fe XXVI. However, the 
seed photon energy and absorption column are unchanged by these 
different models of the iron line structure. Thus the derived 
$N_H$ and $T_{\rm seed}$ are fairly robust.


\begin{table}

\caption{Fits of a multicolour disc blackbody plus thermal 
Comptonization (and its associated reflection spectrum and iron line) 
to the {\it ASCA\/} spectrum. The seed photons for 
Comptonization are assumed to be from the disc, with $T_{\rm seed} = 
T_{\rm in}$. 
}

\label{tab:asca_fits} 
\begin{tabular}{lccc} 
\hline 
Model & $N_H$ (10$^{21}$ cm$^{-2}$) & $kT_{\rm diskbb}$ (keV) & $\chi^2/\nu$\\
\hline
{\sc thcomp} & $6.48_{-0.10}^{+0.09}$ & $0.529_{-0.012}^{+0.012}$ & 209.0/184\\
{\sc compps} & $6.49_{-0.12}^{+0.07}$ & $0.528_{-0.011}^{+0.013}$ & 206.7/184\\
{\sc eqpair} & $6.53_{-0.10}^{+0.10}$ & $0.531_{-0.012}^{+0.012}$ & 207.4/184\\
\hline 
\end{tabular} 
\end{table}


\subsection{OSSE and HEXTE}

The OSSE spectrum extracted from the entire viewing period give 
an excellent fit ($\chi^2/\nu$ = 7.5/19) to a simple power law 
with photon spectral index $\Gamma = 3.09\pm0.04$. There is no 
evidence for any high-energy cutoff in the data with a lower 
limit on an exponential rollover energy of 990~keV.

\begin{figure}
\begin{center} 
\leavevmode 
\epsfxsize=7cm \epsfbox{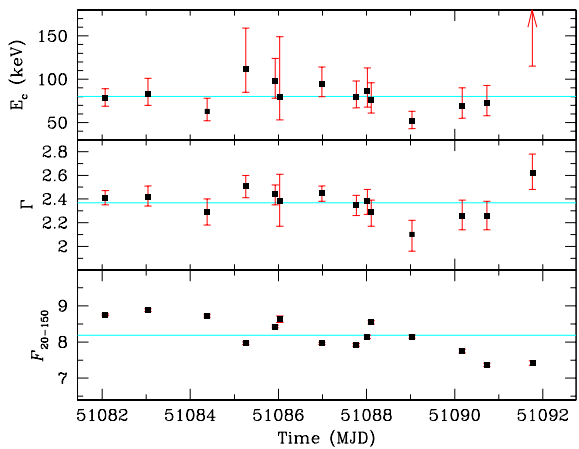} 
\end{center} 

\caption{Variability of XTE J1550--564 in 20--150 keV range 
during OSSE observation. The HEXTE spectra were fitted by the 
cutoff power law and the panels show the e-folding energy, 
$E_c$, photon spectral index, $\Gamma$, and 20--150 keV model 
flux in $10^{-9}$ erg cm$^{-2}$ s$^{-1}$ units. Horizontal lines 
show average values of these parameters. The last observation 
for which only lower limit on $E_c$ exists has been excluded 
from average of $E_c$.}

\label{fig:hxt_var} 
\end{figure}

The HEXTE data are evenly spread throughout the OSSE 
observation, so we can use the better signal to noise of the 
HEXTE instrument to look for spectral variability. The PCA data 
clearly shows that at lower energies there is indeed a small 
spectral change throughout this time, with root-mean-square 
variation in the soft and hard PCA colours of 2.9 and 2.6 per 
cent, respectively (Fig.~\ref{fig:colcol}).  We fit the HEXTE 
spectrum from each {\it RXTE\/} obsid by a cutoff power law and 
Fig.~\ref{fig:hxt_var} shows the resulting values of the 
e-folding energy, spectral index and flux. The spectral shape is 
consistent with remaining constant (with $\Gamma\sim 2.4$ and 
$E_c\sim 80$ keV) despite the 20--150 keV flux decreasing by 
$\sim$ 15 per cent. Hence we are justified in co-adding all the 
HEXTE and OSSE data accumulated over this period to form a 
single spectrum from each instrument. 


\begin{figure}
\begin{center} 
\leavevmode 
\epsfxsize=8cm \epsfbox{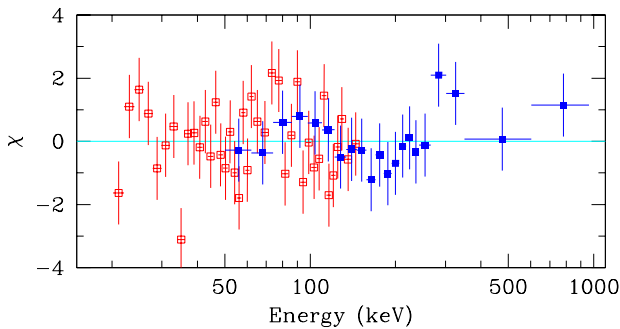} 
\end{center} 

\caption{HEXTE and OSSE residuals. The data were fitted by a thermal 
Comptonization plus power law model. The relative normalization 
between HEXTE and OSSE was free.}

\label{fig:hxt_osse_del} 
\end{figure}


Plainly, the power law and exponential cutoff models have very 
different parameters in the HEXTE and OSSE spectra, so a joint 
fit with this model gives a very poor $\chi^2$. 
Fig.~\ref{fig:hxt_osse_del} shows the residuals of a joint fit 
of the two spectra to a phenomenological model of a power law 
plus thermal Comptonization model. The fit is reasonably good 
($\chi^2/\nu$ = 65.8/54), but more importantly, 
Fig.~\ref{fig:hxt_osse_del} clearly shows that the HEXTE and 
OSSE data are in very good agreement in their region of overlap. 
There are no big cross-calibration problems, or problems with a 
contaminating source in the large OSSE field of view.


\subsection{{\it RXTE\/} and OSSE}
\label{sec:rxte_osse}

Finally, we fit the averaged PCA, HEXTE and OSSE data together with 
the proper Comptonization models. We fix $N_H = 6.5\times10^{21}$ 
cm$^{-2}$ and $kT_{\rm seed} = kT_{\rm diskbb} = 0.53$ keV from the 
{\it ASCA\/} results (Section \ref{sec:asca}). We estimate the inner 
radius from the {\sc diskbb} model, $R_{\rm diskbb}$, assuming 
distance of $D$ = 5.3 kpc, inclination angle of $i = 70^\circ$. We 
correct it for spectral hardening due to scattering, assuming the 
ratio of colour to effective temperature of 1.8 (Shimura \& Takahara 
1995; Merloni, Fabian \& Ross 2000). We also correct it for the inner 
torque-free boundary condition (Gierli{\'n}ski et al.\ 1999).


\begin{figure*}
\begin{center} 
\leavevmode 
\epsfxsize=9.5cm \epsfbox{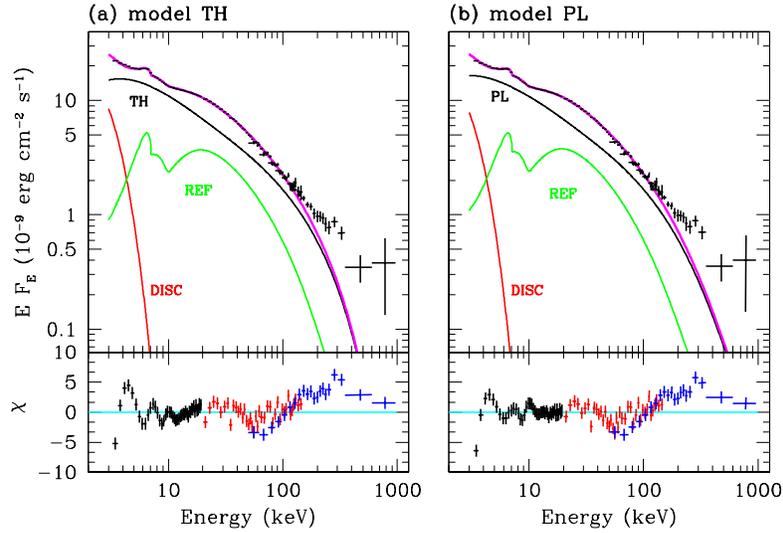} 
\end{center} 

\caption{PCA, HEXTE and OSSE unfolded data, the best-fitting 
models and residuals. ($a$) Comptonization on Maxwellian 
electrons (TH). ($b$) Comptonization on power-law electrons 
(PL). Contribution from the multicolour blackbody disc (DISC) 
and Compton reflection (REF) are also shown.}

\label{fig:spec2} 
\end{figure*}


\subsubsection{Maxwellian and power-law electrons}

Purely thermal Comptonization (a model referred to as TH) cannot 
fit the observed spectrum. A model consisting of the disc 
blackbody, thermal Comptonization (any code described in 
Sec.~\ref{sec:models}) and its reflection/line gives $\chi^2/\nu 
\sim 5$. The best fit is shown in Fig.~\ref{fig:spec2}$a$ and 
clearly shows that the spectral curvature in the lower energy 
data, which requires a fairly low temperature ($kT_e \sim$ 80 
keV, $\tau\sim$ 0.5) thermal Compton component, is strongly 
inconsistent with the higher energy data.  In principle, a very 
high electron temperature ($\gg$ 100 keV) could fit the 
high-energy tail, but to have this together with a fairly steep 
overall spectrum requires a very low optical depth of the 
Comptonizing medium ($\tau < 0.1$). The spectrum from 
Comptonization in such an optically thin plasma consists of 
distinct humps from subsequent scattering orders, which are not 
present in the data. Moreover, the fraction of unscattered disc 
photons would be much larger than seen in the data.

The high-energy tail in the data can be reproduced in the model 
by adding a power law. This gives much better, though still 
statistically unacceptable fit ($\chi^2/\nu$ = 161/93). This is 
because the high-energy data requires fairly hard power law 
($\Gamma \approx 2.3$ from the fit) which extends down to lower 
energies and strongly contributes to the spectrum at soft 
X-rays, distorting it considerably.

Interestingly, a purely non-thermal, power-law electron 
distribution (model PL) cannot fit the data either ($\chi^2/\nu 
\sim 4$; Fig.~\ref{fig:spec2}$b$). This is because the spectrum 
from multiple Compton scatterings on power-law electrons is {\it 
not} a power law. The steep spectrum requires a soft electron 
index (best fit value of $p = 6.3$ from {\sc compps}). This 
means that the average electron energy is low, so photons must 
be scattered many times before they reach high energies ($>$ 100 
keV). In such multiple scatterings they reach the Klein-Nishina 
regime where the cross-section for the scattering decreases. 
This creates a break in the spectrum at a few hundred keV 
(Ghisellini 1989), so underpredicting the higher energy data 
above $\sim 300$ keV. Power-law electrons with very small 
optical depth would be able to produce an unbroken power-law 
photon spectrum on a single scattering. This, however, would 
give too large fraction of unscattered disc photons again.


\begin{figure*}
\begin{center} 
\leavevmode 
\epsfxsize=14cm \epsfbox{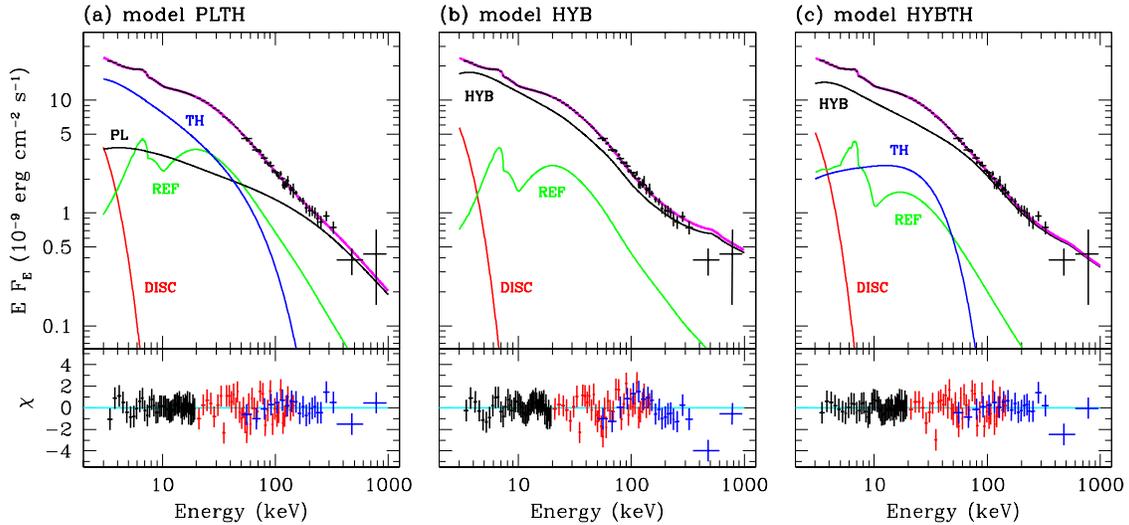} 
\end{center} 

\caption{PCA, HEXTE and OSSE unfolded data, the best-fitting 
models and residuals. ($a$) with two Comptonizing regions: with
power-law (PL) and thermal (TH) electrons. ($b$) Comptonization 
on hybrid (HYB) electron distribution. ($c$) hybrid electrons 
and additional thermal component. Contribution from the 
multicolour blackbody disc (DISC) and Compton reflection (REF) 
are also shown.}

\label{fig:spec3} 
\end{figure*}



\begin{table}

\caption{Fitting results of the model PLTH, consisting of two 
Comptonizations: on thermal (temperature $T_{\rm th}$ and 
optical depth $\tau_{\rm th}$) and power-law (index $p$ and 
optical depth $\tau_{\rm pl}$) electrons. Values in parentheses 
refer to the parameters fixed during the fit. The unfolded data 
and model of this fit are shown in Fig.~\ref{fig:spec3}$a$.}

\label{tab:compps} 
\begin{center}
\begin{tabular}{ll} 
\hline
& PLTH\\
\hline
$R_{\rm diskbb}$ (km)    & $150_{-13}^{+15}$\\
$kT_{\rm th}$ (keV)      & $27_{-4}^{+5}$\\
$\tau_{\rm th}$          & $1.6_{-0.2}^{+0.3}$\\
$\tau_{\rm pl}$          & $0.29_{-0.07}^{+0.08}$\\                 
$p$                      & $4.5_{-0.3}^{+0.3}$\\
$\Omega/2\pi$            & $1.22_{-0.13}^{+0.08}$\\
$\xi$ (erg cm s$^{-1}$)  & $42_{-7}^{+14}$\\
$R_{\rm ref}$ ($R_g$)    & (6)\\
$E_{\rm line}$ (keV)     & $5.60_{-0.06}^{+0.06}$\\
EW (eV)                  & 220\\
$\chi^2/\nu$             & 65.9/94\\
\hline 
\end{tabular} 
\end{center}
\end{table}


Plainly, the data require a more complex continuum. Now we use a 
model PLTH, consisting of two Comptonizing regions (both computed 
with {\sc compps}): one with a Maxwellian (thermal) electron 
distribution, the other with power-law electrons. Both these 
Comptonize the seed photons from the disc independently. We obtain a 
very good fit with $\chi^2/\nu = 65.9/94$ (see Table 
\ref{tab:compps}). The spectrum at lower energies is dominated by the 
thermal component with temperature about 30 keV, while the 
high-energy tail is created by up-scattering seed photons on 
optically thin ($\tau \approx 0.3$), steep power-law ($p \approx 
4.5$) electrons, where the break from the Klein-Nishina is masked by 
the thermal Compton component (Fig.~\ref{fig:spec3}$a$).

The requirement for both thermal and non-thermal electrons does not 
simply rest on the (rather weak) detection of the source at the 
highest energies in OSSE ($3.7\sigma$ and $1.5\sigma$ for channels 
350--600 keV and 600--900 keV, respectively). Instead it is driven by 
the overall spectral shape between $\sim$ 100 and $\sim$ 300 keV 
where the OSSE data have excellent statistics.  Even the spectral 
fits with the OSSE data above 350 keV excluded strongly rule out 
purely thermal or purely power-law Comptonization models, as these 
both produce a much steeper high-energy cutoff than observed.

\subsubsection{Hybrid electrons}

Alternatively, the complex continuum could be produced by a single
Comptonizing plasma, where non-thermal electrons partially thermalize
through Coulomb collisions, creating a hybrid distribution: Maxwellian
at lower energies and power-law at higher energies.  We use {\sc
eqpair} to compute this distribution self-consistently, together with
its reflection/line emission and seed photons from the disc (model
HYB). 

This hybrid model gives a somewhat worse fit ($\chi^2/\nu = 
100.6/94$, see Table \ref{tab:eqpair} and 
Fig.~\ref{fig:spec3}$b$) than the two Comptonizing regions model 
PLTH ($\chi^2/\nu =63.4/92$), although it is still statistically 
an adequate description of the data. The reason the fit is 
different is that Comptonization from two separate electron 
distributions, where one is thermal and the other non-thermal, 
gives a subtly different spectrum to that of a single hybrid 
electron distribution. The sum of a thermal and power-law 
distribution has the power-law electrons also contributing to 
the spectrum at low energies, whereas in the hybrid electron 
distribution there are only the thermal electrons at low 
energies. While in principle this could give us a way to 
distinguish between the two models, in practice the situation is 
much less clear due to uncertainties in the reflected spectrum 
(see below). 

Error ranges are hard to obtain as many of the parameters are 
correlated, so give multiple local minima in $\chi^2$. In 
particular, the soft compactness, \ls, is anti-correlated with 
the fraction of power in the non-thermal injection, \lnh. This 
is why in the model HYB we fixed \ls\/ = 10. The soft 
compactness determines the strength of Compton cooling relative 
to Coulomb cooling, so a larger fraction of the non-thermal 
electrons thermalize for small \ls. The effects of this can be 
offset by putting less power into the thermal heating of the 
electrons i.e. by increasing \lnh. A model HYBN (see Table 
\ref{tab:eqpair}) with all the hard power in the non-thermal 
injection (fixed \lnh\/ = 1, but \ls\/ left free) is 
statistically acceptable ($\Delta\chi^2=+1.7$ compared to model 
HYB). This model has \ls\/ $\approx$ 0.8, which is a firm lower 
limit on the soft compactness. 

An upper limit on the soft compactness can also be found from the
spectrum.  Since the hard-to-soft compactness ratio is very 
well determined by the spectral shape, \lhs\/ $\approx$ 1, large 
\ls\/ requires large \lh. But there is an upper limit on the 
hard compactness, since large \lh\/ gives rise to a strong
annihilation line which is not present in the data. Thus, we can 
roughly constrain the soft compactness, 1 $\la$ \ls\/ $\la$ 20 
and the total compactness, 2 $\la \ell \la$ 40.

The contribution of electron-positron pairs to the Comptonizing 
plasma is negligible, as the best-fitting model yields $\tau_{\rm 
pair} \sim 10^{-3}$. The plasma would be pair-dominated ($\tau_{\rm 
pair} > \tau_{\rm ion}$) only for extremely high compactness, $\ell 
\ga 3000$, which is ruled-out by the data ($\chi^2/\nu \sim 128/94$).


\begin{table}

\caption{Fitting results with a hybrid thermal/non-thermal model. 
HYB is the hybrid model with fixed \ls\/ = 10 and free \lnth. HYBN is 
the hybrid model with free \ls\/ and fixed \lnth\/ = 1 (no 
thermal heating). HYBTH stands for the hybrid model with 
additional component with thermal electrons. Values in 
parentheses refer to the parameters fixed during the fit. The 
equilibrium temperature, $T_e$, was calculated self-consistently 
and was not a fit parameter.}

\label{tab:eqpair}
\begin{center}
\begin{tabular}{llll} 
\hline
& HYB & HYBN & HYBTH\\
\hline
$R_{\rm diskbb}$ (km)    & $197^{+7}_{-7}$         & $200^{+6}_{-6}$        & $186_{-10}^{+10}$ \\
$kT_{\rm th}$ (keV)      & --                      & --                     & $7.7_{-0.6}^{+0.5}$\\
$\tau_{\rm th}$          & --                      & --                     & $6.4_{-1.0}^{+0.8}$ \\
\ls                      & (10)                    & $0.81_{-0.24}^{+0.17}$ & (10) \\
\lnh                     & $0.59_{-0.04}^{+0.03}$  & (1.00)                 &(0.59) \\
\lhs                     & $1.04_{-0.04}^{+0.04}$  & $0.90_{-0.03}^{+0.04}$ & $0.94_{-0.03}^{+0.04}$ \\
$kT_e$ (keV)             & 12                      & 20                     & 20\\
$\tau_{\rm ion}$         & $2.78_{-0.06}^{+0.07}$  & $1.77_{-0.15}^{+0.06}$ & $1.76_{-0.08}^{+0.07}$ \\ 
$\Gamma_{\rm inj}$       & $3.19_{-0.09}^{+0.06}$  & $3.22_{-0.12}^{+0.07}$ & $3.34_{-0.08}^{+0.08}$ \\
$\Omega/2\pi$            & $0.76_{-0.09}^{+0.07}$  & $1.17_{-0.09}^{+0.05}$ & $0.32_{-0.04}^{+0.12}$ \\ 
$\xi$  (erg cm s$^{-1}$) & $190_{-50}^{+110}$      & $100_{-10}^{+15}$      & $8000_{-6000}^{+17000}$ \\
$R_{\rm ref}$ ($R_g$)    & (10)                    & (10)                   & (30) \\
$E_{\rm line}$ (keV)     & $5.89_{-0.06}^{+0.06}$  & $5.84_{-0.04}^{+0.07}$ & $6.23_{-0.06}^{+0.07}$\\
$\chi^2/\nu$             & 100.6/94                & 102.3/94               & 63.4/92\\
\hline 
\end{tabular} 
\end{center}
\end{table}


The Compton reflection requires a substantial amount of relativistic 
smearing. We fix the inner disc radius at $R_{\rm ref} = 10R_g$ 
during the fit, but values higher that $\sim$ 15$R_g$ are ruled-out. 
The best-fitting model give unphysical rest-frame iron line energy of 
5.6 keV. This probably indicates that the Compton reflection model 
we use here is inadequate, as already inferred from the 
{\it ASCA\/} fits (Section \ref{sec:asca}).

These fits can be dramatically improved ($\chi^2/\nu = 63.4/92$, 
model HYBTH) by including an additional, optically thick, 
thermal Comptonization component (for which we use {\sc thcomp} 
code).  The residuals in Figs.~\ref{fig:spec3}$b$ and 
\ref{fig:spec3}$c$ show that the improvement affects the whole 
spectral range, though in particular the features below $\sim$ 
10 keV are substantially smoothed. The model is detailed in 
Table \ref{tab:eqpair}, but again the complex correlations 
between parameters mean that again the errors are only 
representative as we fixed \lnh\/ = 0.59 (from model HYB) and 
$R_{\rm ref} = 30R_g$.

The hot plasma properties are insensitive to the presence of the 
additional thermal Compton 
component. Both HYB and HYBTH models predict \lhs\/ $\approx$ 
1 and a significant fraction of the hard power from non-thermal 
electrons (\lnh\/ $\ga$ 0.5). They both rule out pair-dominated 
plasma. The main difference is in the optical depth of the 
Comptonizing medium, which decreased from $\sim$2.8 to $\sim$1.8 when 
{\sc thcomp} was added.

The extrapolated unabsorbed bolometric flux of XTE~J1550--564 from 
both models is $1.1\times10^{-7}$ erg cm$^{-2}$ s$^{-1}$. At a 
distance of 5.3 kpc, for a 10M$_\odot$ black hole, this corresponds 
to about $0.3L_{\rm Edd}$.

\section{High and very high X-ray spectral states}
\label{sec:states}


\begin{figure}
\begin{center} 
\leavevmode 
\epsfxsize=7cm \epsfbox{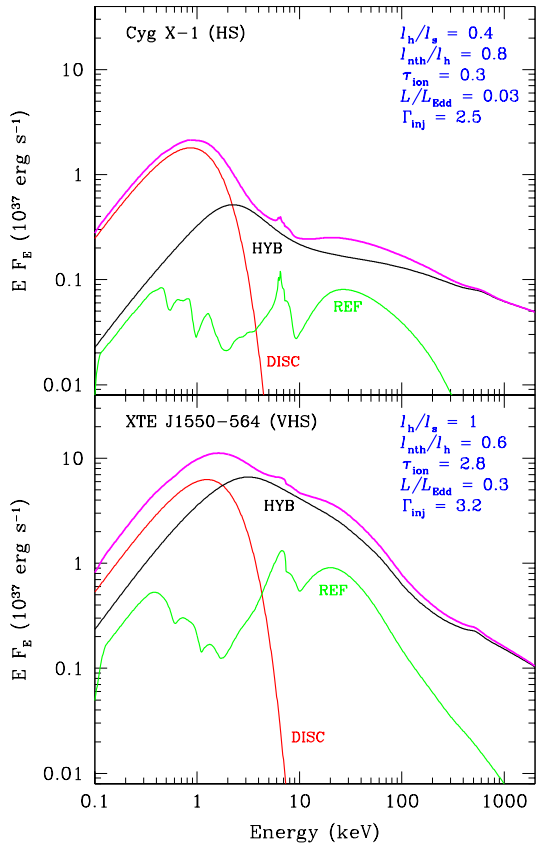} 
\end{center} 

\caption{Unabsorbed model components for Cyg X-1 in the high state 
(from Gierli{\'n}ski et al.~1997) and XTE J1550--564 in the very 
high state (this paper, model HYB). The approximate values of some of 
the best-fitting {\sc eqpair} parameters are given. To convert fluxes 
into luminosities we assumed distances of 2 kpc for Cyg X-1 and 5.3 
kpc for XTE J1550--564. For Eddington luminosities we assumed 
10M$_\odot$ black hole mass for both sources.}

\label{fig:hs_vhs} 
\end{figure}


As we have outlined in the introduction, there are at least two 
distinct soft spectral states of black hole binaries: high and 
very high. We attribute our observation of XTE J1550--564 to the 
very high state. How does it compare with other sources, in both 
soft states? Fig.~\ref{fig:hs_vhs} shows a comparison of the 
high- and very high-state model spectra. The high-state spectrum 
is from Cyg X-1 (Gierli{\'n}ski et al.~1999) while the very 
high-state one is from XTE J1550--564 (this paper). The same 
model is used for both, with a hybrid electron distribution 
calculated using the {\sc eqpair} code (Coppi 1999).

The main difference between the high and very high states seems 
to be the amount of Comptonization of the disc spectrum.  In the 
high state many of the disc photons escape without being 
strongly Comptonized, therefore the Comptonizing plasma either 
covers only a small fraction of the disc and/or has low optical 
depth. In the very high state most of the disc emission is 
scattered to substantially higher energies, so the Comptonizing 
medium must have a substantial optical depth and cover most of 
the disc. 

However, we caution that the amount of Comptonization is not 
always straightforward to derive from the spectra. For example, 
Zdziarski et al.~(2001) fit two soft state {\it RXTE}/OSSE 
spectra from GRS 1915+105. The `ultrasoft' high-state spectrum 
(VP 813) has a strong, dominating soft component and a weak 
high-energy tail. It is similar to the Cyg X-1 spectrum from 
Fig.~\ref{fig:hs_vhs}, but the temperature of the soft component 
and its ratio of power to the hard tail is larger. The second 
observation analyzed by Zdziarski et al.~(2001) is a typical 
very high-state spectrum (VP 619), with comparable distribution 
of power between the disc and the tail, similar to the XTE 
J1550--564 spectrum analysed here (see fig.~2 in Zdziarski et 
al.~2001). However, the detailed spectral modelling with {\sc 
eqpair} suggests a rather different interpretation of this high 
state to that outlined above. What looks like a disc in the 
high-state spectrum of GRS 1915+105 is fit by Comptonization of 
the disc in optically thick, low-temperature plasma, very 
different to the high state of Cyg X-1. However, it seems more 
likely that this simply reflects the fact that the disc spectral 
models do not adequately describe the observed disc emission. 
The dominating soft component in GRS 1915+105 probably comes 
from the disc itself, while an optically thin, hot plasma is 
responsible for the weak hard tail. Thus we conclude that there 
is probably an increase in the covering fraction and/or optical 
depth of the Comptonizing plasma which drives the source 
evolution from the high to very high states.

Another difference between the high and very high states is a 
change in shape of the non-thermal tail. The power-law index of 
the injected electron distribution is steeper in the very high 
state than in the high state, as is also seen in GRS 1915+105 
(Zdziarski et al.~2001). Thus it seems that there are also 
changes in the non-thermal electron acceleration associated with 
the source moving from high to very high state.


\section{Discussion and conclusions}
\label{sec:discussion}

During its 1998/1999 outburst the black hole candidate XTE J1550--564 
showed a variety of X/$\gamma$-ray spectral states. In this paper we 
have analyzed its broadband spectrum during the very high state. This 
spectral state is characterised by the presence of the soft, 
high-energy tail extending without any apparent break above $\sim$ 
300 keV. Such a tail differs from the low/hard state tail both in 
having a significantly softer spectral index and in the lack of 
high-energy cutoff (Grove et al.~1998). Phenomenologically, the OSSE 
spectra of these tails can be described as a power law with spectral 
index $\Gamma \sim$ 2.4--3.1 (Grove et al.~1998). In this observation 
$\Gamma = 3.09\pm0.04$. We have shown that the broadband 
X/$\gamma$-ray spectrum of XTE J1550--564 can be well fit by a model 
in which seed photons from the disc are Compton up-scattered by both 
thermal and non-thermal electrons. Comptonization on purely thermal 
or purely power-law electrons can be ruled out.

The thermal and non-thermal electron distributions may be distinct, or
they may instead form a single hybrid thermal/non-thermal
distribution. This is reflected in the modelling of the XTE J1550--564
spectrum, which we have fitted by separate thermal plus non-thermal
(model PLTH) and hybrid (model HYB) Comptonizations.  However, even if
there is a spatially separate non-thermal electron population, these
electrons will interact via Coulomb collisions, creating a thermal
low-energy extension to the power-law distribution. Thus, we expect a
hybrid distribution of Comptonizing electrons even from a completely
non-thermal electron acceleration mechanism. Therefore, the PLTH model
seems to be less physically motivated, and a single hybrid plasma
(HYB) would be our model of choice as the simplest, physical
description of the data. However, spectral fitting gives the opposite
result, with the less physical model (PLTH) being strongly
statistically preferred.

Since we expect that the non-thermal electrons form a hybrid plasma,
the next step in complexity would be two separate electron
distributions, with one thermal and one hybrid (rather than one
thermal and one power law). This model (HYBTH) provides an excellent
fit to the data, statistically similar to PLTH but without using an
unphysical power law electron distribution. However, closer
examination of the Maxwellian component in HYBTH (component denoted as
TH in Fig.~\ref{fig:spec3}$c$) reveals its suspicious similarity in
spectral shape to the Compton reflected continuum. Perhaps it is not
due to emission from some thermal, optically thick plasma at
all. Perhaps it simply accounts for inaccuracies of the reflection
model. In particular, the reflection model used here overestimates the
depth of the iron edge for ionized material as it does not include
Compton upscattering (Ross, Fabian \& Young 1999). While the single
hybrid plasma (HYB) is {\em statistically\/} ruled out when compared
to the two-component solution (HYBTH), better reflection models are
required before this can be a robust conclusion.

Whatever is the physical reality behind this additional thermal
Compton component, the properties of the hybrid plasma are established
quite robustly and do not change much between HYB and HYBTH. The ratio
of soft to hard compactness is very well defined by the shape of the
spectrum (\lhs\/ $\approx$ 1), while the fraction of the power in
non-thermal electrons is \lnh\/ $\ga$ 0.5. Once again, the presence of
non-thermal electrons is required to form the high energy tail
observed beyond 300~keV. But where do they come from?

One obvious mechanism for such high-energy emission is from a 
jet, as in the blazars and radio-loud quasars (see e.g. 
Ghisellini et al.~1998). A jet is present in XTE J1550--564 
indeed at this time, although its radio emission is declining 
rapidly throughout the OSSE observation (Wu et al.~2002), 
so its link with the X/$\gamma$-ray emission is doubtful. 
Moreover, the fact that a similar high-energy tail is seen in 
the high state as well as in the very high state immediately 
rules out a direct association with the jet as the radio 
emission is strongly quenched in the high state (Corbel et al. 
2000).

An alternative mechanism of bulk motion of in-falling material 
near the black hole horizon was proposed by Chakrabarty \& 
Titarchuk (1995), but the free fall results in fairly low 
electron energies, with Lorentz factors $\gamma \la 1.4$. These 
electrons are unable to up-scatter photons past $\sim$ 100--300 
keV, resulting in a cutoff in the power-law spectrum (Laurent \& 
Titarchuk 1999; Zdziarski et al.~2001). Since such a cutoff is 
not seen in the XTE J1550--564 data, we find the bulk motion 
Comptonization to be an unlikely explanation of the observed 
emission.

This leaves magnetic reconnection above the disc as the only 
known possible source of hard X-ray dissipation at high mass 
accretion rates. The accretion disc viscosity is now known to be 
linked to a magnetic dynamo, so magnetic buoyancy can lead to 
some fraction of the reconnection events occurring in a corona 
rather than deep inside the disc (see e.g.\ Hawley \& Balbus 
2002), accelerating electrons to high energies (di Matteo 1998). 
We speculate that the high and very high state spectra are 
continuous rather than two separate spectral states, linked by 
an increase in the covering fraction of this magnetic corona 
over the disc.

Concluding our paper we summarise our main results:\begin{itemize}
\item The broad-band X/$\gamma$-ray spectrum of XTE J1550--564 
in the very high state can be well described by inverse Compton 
emission from hybrid, thermal/non-thermal plasma.
\item Comptonization on purely thermal or purely power-law electrons 
cannot explain the observed spectrum.
\item The high-energy power-law tail observed here is similar in 
other black hole binaries in the high and very high states, showing
that it is a generic property of high mass accretion rate black
holes. 
\end{itemize}

\section*{Acknowledgements}

This research has been supported in part by the Polish KBN grant 
2P03D00514. We thank John Tomsick for providing the computer 
code for the PCA pile-up correction and Aya Kubota for useful 
discussions.


\end{document}